\documentclass[letterpaper]{article} % DO NOT CHANGE THIS
\usepackage{aaai25}  % DO NOT CHANGE THIS
\usepackage{times}  % DO NOT CHANGE THIS
\usepackage{helvet}  % DO NOT CHANGE THIS
\usepackage{courier}  % DO NOT CHANGE THIS
\usepackage[hyphens]{url}  % DO NOT CHANGE THIS
\usepackage{graphicx} % DO NOT CHANGE THIS
\usepackage{natbib}  % DO NOT CHANGE THIS AND DO NOT ADD ANY OPTIONS TO IT
\usepackage{caption} % DO NOT CHANGE THIS AND DO NOT ADD ANY OPTIONS TO IT
\usepackage{algorithm}
\usepackage{algorithmic}
\usepackage{pdfpages}

\usepackage{newfloat}
\usepackage{listings}
\DeclareCaptionStyle{ruled}{labelfont=normalfont,labelsep=colon,strut=off} % DO NOT CHANGE THIS
\floatstyle{ruled}
\newfloat{listing}{tb}{lst}{}
\floatname{listing}{Listing}
\usepackage{xspace}

\newcommand*{\eg}{e.g.,\xspace}

\title{Using Case Studies to Teach Responsible AI to Industry Practitioners}

\author{
   Julia Stoyanovich,
    Rodrigo Kreis de Paula,
    Armanda Lewis,
    Chloe Zheng
}
\affiliations{
    New York University, New York, NY, USA\\
    \{stoyanovich,rodrigo.kreis,al861,cz1300\}@nyu.edu 
}

\begin{document}

\maketitle

\begin{abstract}

Responsible AI (RAI) encompasses the science and practice of ensuring that AI design, development, and use are socially sustainable--—maximizing the benefits of technology while mitigating its risks. Industry practitioners play a crucial role in achieving the objectives of RAI, yet there is a persistent a shortage of consolidated educational resources and effective methods for teaching RAI to practitioners.

In this paper, we present a stakeholder-first educational approach using interactive case studies to foster organizational and practitioner-level engagement and enhance learning about RAI. We detail our partnership with Meta, a global technology company, to co-develop and deliver RAI workshops to a diverse company audience. Assessment results show that participants found the workshops engaging and reported an improved understanding of RAI principles, along with increased motivation to apply them in their work.

\end{abstract}

\section{Introduction}
\label{sec:intro}

Responsible AI (RAI) encompasses the science and practice of ensuring that AI design, development, and use are socially sustainable--—maximizing the benefits of this technology while mitigating its risks. Widespread recognition of the importance of RAI has led to recent legislative and regulatory decisions, and high-level directives.  Notable examples include the recent U.S. Executive Order on Safe, Secure, and Trustworthy Artificial Intelligence (EO 14110)~\cite{biden_jr_executive_2023}, which commits to ``addressing safe, responsible, and non-discriminatory uses of AI,'' and the European Union's adoption of the Artificial Intelligence Act~\cite{ european_union_artificial_2023}. 

Stakeholders, including technical developers, designers, end-users, others impacted by AI, and society at large, have distinct priorities.  For this reason, deep engagement with the tensions between differing stakeholder perspectives is necessary to build and deploy AI systems responsibly.  Industry practitioners play a decisive role in our collective ability to achieve the goals of RAI, making it essential for them to collaborate with academic institutions to integrate RAI advances into applications and services. We describe such a collaboration in this paper.   

We are members of the Center for Responsible AI at New York University (NYU R/AI), an academic institution dedicated to advancing RAI principles through research, education, and collaboration. We partnered with Meta, a global technology company, which had initiated internal advocacy efforts to promote RAI and recognized the importance of leadership-level support to embed these principles into its organizational culture. With input from Meta staff, we developed  RAI training materials.  We then offered workshops to equip legal, managerial, and technical professionals within the company with the knowledge and skills needed to integrate RAI principles into the design, development, and deployment of AI systems.

The driving questions behind our work are: ``How can we engage organizational partners to better integrate RAI as a core component of their institutional goals?'' and  ``What strategies facilitate diverse practitioners---across design, legal, and technical teams---to learn about and incorporate RAI principles in their daily work?''  To answer these questions, we followed a stakeholder-first approach to training~\cite{dominguez_2023}, using \emph{interactive case studies}  to achieve organizational and practitioner-level engagement, promote inter-role collaboration, encourage a multi-perspective approach, and advance RAI learning in a practical environment.  By grounding learning in practical, real-world scenarios, this approach not only deepens understanding of RAI principles but also fosters inter-role collaboration, making it more likely that these principles are embedded into organizational processes and decision-making.

This collaboration posed several challenges, including lack of access to information about the company's proprietary algorithms and systems during case study development. Additionally, organizational changes at Meta during the collaboration led to team restructuring, complicating efforts to achieve collective buy-in. Nonetheless, our qualitative and quantitative analysis of the workshops indicate that participants--- many with limited prior knowledge---found the training engaging and reported a positive shift in their understanding and motivation to apply RAI concepts to their work. The use of case studies relevant to participants' day-to-day  activities was particularly well-received. 

\section{Related Work} 

The concept of the stakeholder---``any group or individual who can affect or is affected by the achievement of the organization's objectives''~\cite{freeman_2010} 
---has long been a focus in organizational research and is increasingly relevant in AI contexts. Scholars have noted the importance of incorporating multi-stakeholder perspectives, including decision-makers, companies, shareholders, government and regulatory bodies, technologists, end-users, and society at large, at all stages of algorithmic system development \cite{gungor_creating_2020,lima_responsible_2020,nabavi_leverage_2023}. In examining stakeholder influence within the AI lifecycle, ~\citet{miller_stakeholder_2022} extends the standard classification of~\citet{mitchell_toward_1997}---power, legitimacy, and urgency---by adding harm as another attribute,  enabling the identification of additional groups potentially affected during the development and operating stages of an AI project.  

Additionally, research has looked at integrating multiple stakeholders into the technical aspects of design and deployment processes, particularly targeting AI system designers, engineers, and researchers.  For example,~\citet{bell_think_2023} argue that technologists, equipped with both expertise and organizational influence, are well-positioned to drive meaningful change. They propose a four-tier hierarchy to guide the design and deployment of transparent automated decision systems.  Within this framework,  practitioners consider diverse stakeholders---such as affected individuals, policymakers, and other technologists---and use these insights to formulate appropriate goals, purposes, and methods.  Further, \citet{abdollahpouri_multi-stakeholder_2019},\citet{abdollahpouri_multistakeholder_2020}, and~\citet{bell_possibility_2023} demonstrate that fair machine learning models can be effectively developed through a stakeholder-driven approach, which integrates the priorities of practitioners with the needs of affected individuals, policymakers, and other technologists, all while maintaining minimal trade-offs in accuracy.

While we recognize the importance of individual stakeholders, we also emphasize the role of organizational representatives in advancing AI.  These representative can drive adoption of RAI by elevating responsibility within the organization and signaling its importance to both internal and external stakeholders.

\begin{figure}[t!]
    \centering
    \includegraphics[width=\linewidth]{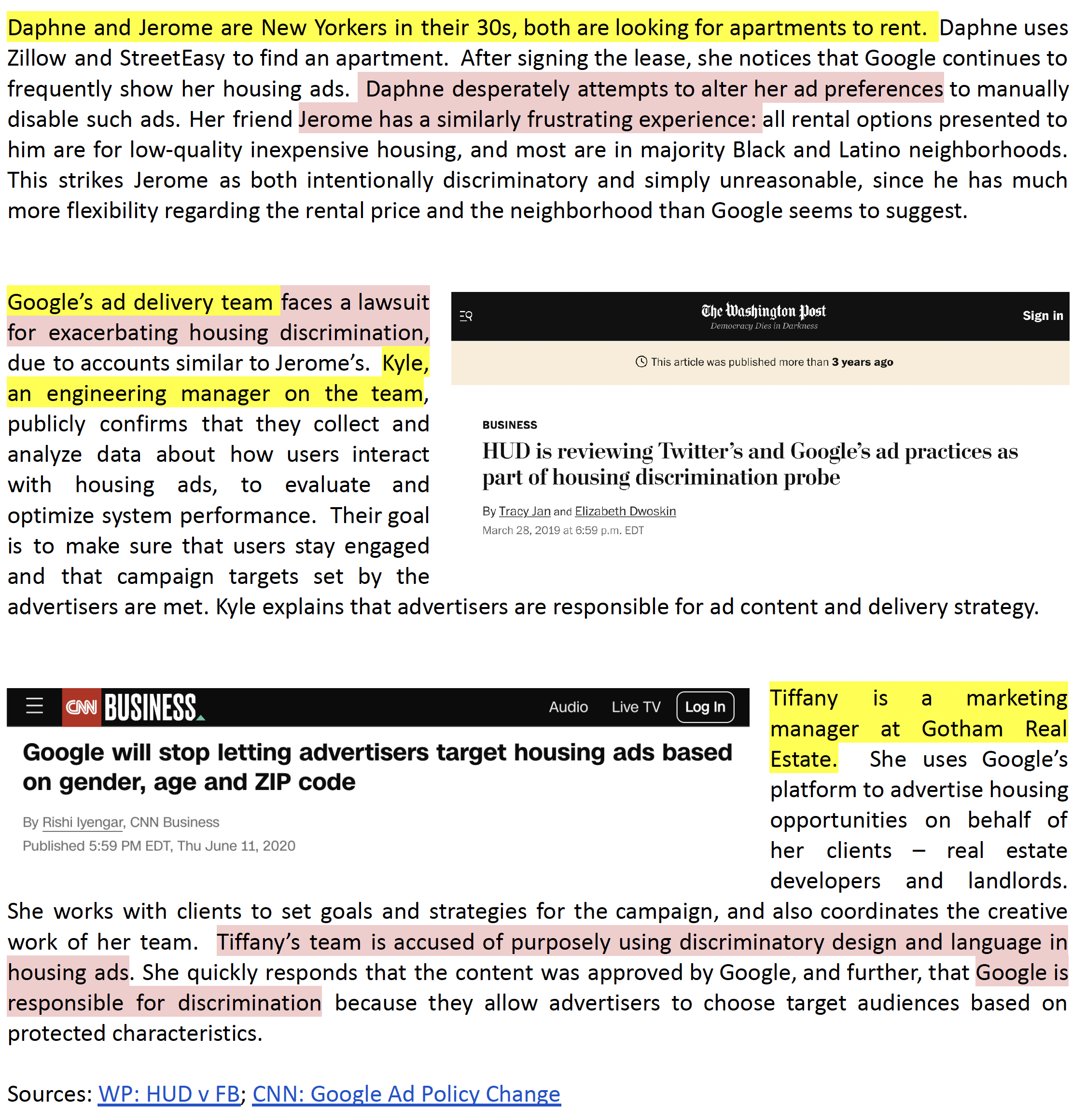}
    \caption{Housing ads case study: negative sentiment hand-out.   Yellow highlights emphasize stakeholders, and red highlights emphasize the negative sentiment.  Highlights were not part of the handout used during the workshops.  See Supplementary Materials for a complete set of handouts.}
    \label{fig:housing_handout_negative}
\end{figure}

Our process includes two essential levels of engagement to advance practitioners’ understanding of RAI principles.  The first involves engaging with \emph{organizational partners} and higher-level managerial representatives, leveraging their active participation and institutional buy-in to shape the practitioner experience, which centers on instructive RAI case studies \cite{freeman_stakeholder_2017,shah_stakeholder_2022}.  The second focuses on engagement with \emph{practitioners} responsible for technical and non-technical deliverables.  This approach involves direct instruction in RAI principles and aligns with educational policy literature, which reports direct learning interventions as essential for sustainable institutional change \cite{goldsmith_why_2017,henry_educational_2013}.  We integrate pedagogical best practices by designing active learning experiences that enable practitioners to learn by doing and construct knowledge based on their roles, perspectives, and interests, fostering bottom-up engagement around integrating RAI principles \cite{lewis_teaching_2022,sawyer_cambridge_2022,wenger_communities_1998}.  \emph{Our primary method for fostering active learning among practitioners is the implementation of thoughtfully crafted case studies.}
\section{Engaging the Organizational Partner}
\label{sec:org}

We are a team of academic researchers and educators from NYU R/AI\footnote{\url{https://r-ai.co}},
a U.S.-based academic center, referred to here as the \emph{academic partner}. We conducted this work in close collaboration with Meta, a major tech corporation, referred to as the \emph{organizational partner}. Like many other technology companies today, Meta has a RAI team that implements and disseminates RAI practices within the organization. This team initiated our collaboration to develop and iteratively refine training materials on RAI for the organization's staff and to offer workshops using these materials. Our collaboration began in Fall 2021 (when the company still operated as Facebook) and concluded in Spring 2023.  During this time, the \emph{organizational partner} underwent structural changes, which led to several shifts in the project's ``ownership'' within the company.

In Fall 2021, we collaboratively designed a \emph{pilot workshop} titled ``Demystifying Responsible AI,'' offered over four 60-minute sessions in Winter 2022. While several pilot participants were part of the company's RAI team or worked closely with it, they did not represent the majority.   The pilot workshop covered four main themes: (1) What is responsible AI? (2) Transparency and interpretability; (3) Algorithmic fairness; and (4) A lifecycle view of responsible AI.  Each session followed a consistent structure, starting with a brief, discussion-based warm-up activity, followed by a two-part presentation from the lead instructor. Between the presentation segments, there was a 10-minute discussion, with an additional discussion at the end of the session.  Instructional materials included a rich set of examples and featured several case studies, albeit briefly presented.  Participants also received optional reading, consisting of general-audience and scientific articles on RAI. 

We collected informal feedback from pilot workshop participants to inform the next round of iterative development.  Overall, the feedback was positive, while listing areas for improvement. \emph{First}, several participants noted that they enjoyed the examples and case studies presented during each session.  They suggested---including agreement from the organizational partners---that additional case studies should be incorporated into future iterations. Further, participants emphasized the importance of aligning case studies with their day-to-day professional activities. \emph{Second}, several participants found discussions and Q\&A to be the most engaging aspect of the workshop. They suggested making the workshop more interactive by reducing instructor-led presentation time and extending group discussion time. \emph{Third}, although about 20 participants joined the first session, there was substantial attrition by the fourth session, possibly due to work-related demands that may have reduced participants' availability to attend all workshop sessions.  To address this and reduce the human resources cost on the company, the \emph{organizational partner} suggested condensing the workshop into fewer sessions.

We incorporated these suggestions and immediately commenced the difficult work of identifying and developing case studies for the next iteration of the workshop, as detailed in next section.  We also redesigned the workshop's structure in line with suggestions from the \emph{organizational partner} and pilot workshop participants. The second iteration, offered in early Spring 2023, was titled ``What is Responsible AI and how does it apply to your work at Meta?'' to emphasize the connection between the content and participants' day-to-day activities. We will describe the structure and content of the workshop under Workshop Implementation.

\section{Case Studies for Teaching RAI}
\label{sec:casestudies}

\begin{figure*}
    \centering
    \includegraphics[width=\linewidth]{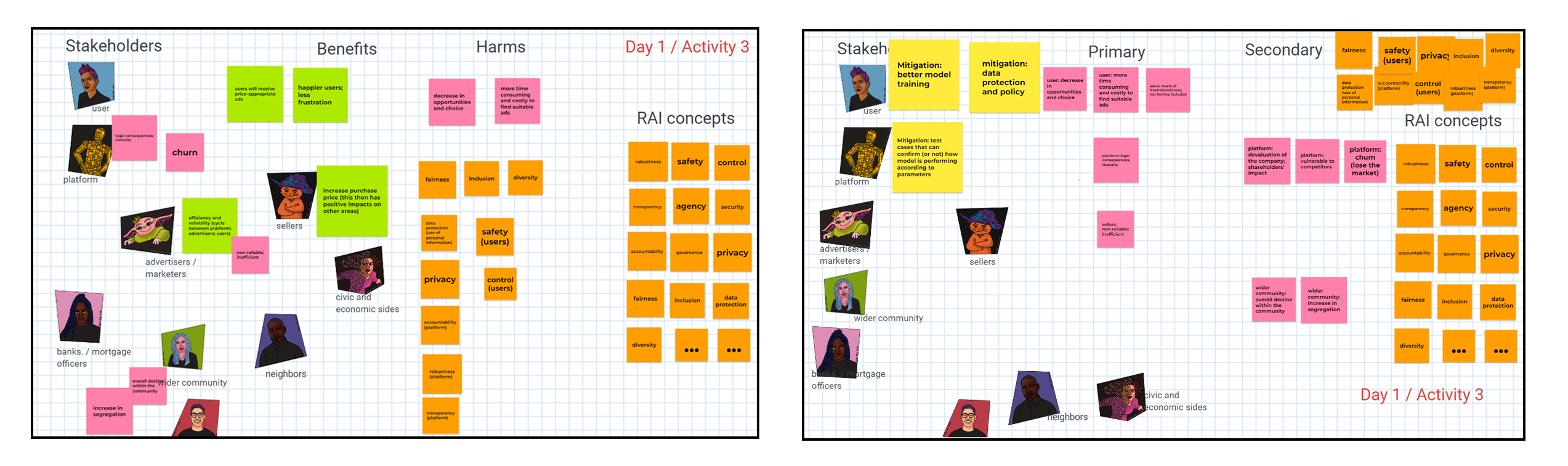}
    \caption{Jamboards for the housing ads case study: benefits, harms, and mitigation strategies.}
    \label{fig:jamboard_day1}
\end{figure*}

Creating and deploying effective case studies is pivotal in the pedagogical pursuit of raising RAI awareness and knowledge among practitioners. Case studies offer a pragmatic and investigative approach to understanding the complexities, challenges, and ethical dimensions of AI architecting, building, and implementation \cite{turner_case_2014}. More generally,~\citet{kreber_learning_2001} argues that case studies provide essential experiential learning opportunities, prompting reflection, reconceptualization, and applied use of learned materials when implemented effectively. The literature on teaching AI and AI ethics~\cite{hishiyama_educational_2022, khan_ethics_2022, laine_ethics-based_2024} highlights that case studies help learners ground ethical principles in realistic scenarios and reveal tensions. 

We follow this approach and use high-quality case studies based on current AI systems, platforms, and scenarios to prompt practitioners to \emph{identify stakeholder benefits and harms, evaluate trade-offs, and reconcile tensions}. Case studies are also effective at fostering organizational improvement. Exploring real-world scenarios provides practitioners with a sandbox to refine processes, policies, and overlooked ethical issues, including competing interests and complex trade-offs \cite{garrett_more_2020,kazim_high-level_2021,dominguez_2023}. This section explores our most relevant findings on designing case studies for teaching RAI and presents the case studies developed for the 2023 workshops.

\paragraph{Case study selection.}
Creating a case study that resonates with the audience begins with carefully  \emph{considering the individuals} who will be studying it. Practitioners come from diverse backgrounds with varying skills and professional goals, so understanding their knowledge and objectives is vital to ensuring the case study meets their educational needs. Alongside focusing on the audience, we prioritized \emph{choosing cases rich in detail and complexity}. These cases present a range of challenges and ethical issues related to AI, fostering immersive learning and providing practitioners with a broad view of its landscape. 
We also emphasized the value of \emph{cases with diverse stakeholders, whose differing perspectives can reveal conflicting interests and tensions}.  

Finally, the organizational partner encouraged us to \emph{choose cases that were directly relevant to their organization and}  \emph{participants' day-to-day activities}.  Satisfying this requirement proved difficult, as we lacked access to 
internal information about Meta's products and services.  Additionally, we aimed to develop case studies that could be shared publicly to support broader RAI training efforts.  For these reasons, we selected case studies with sufficient publicly available information that were also broadly relevant to Meta's application domain and industry but that do not center around any of their strategic products or services.

Having selected the case studies, we proceeded to document each one comprehensively.

\paragraph{Case study documentation.}  We comprehensively documented each case study using the following format.

\emph{Overview:} A succinct, informative introduction of the AI system, setting the stage for subsequent exploration, offering context, and framing key aspects to be examined. 

\emph{Background and context of use:} Insights into the broader background and context of the AI system, including its implementation timeline and challenges, to help practitioners grasp its real-world significance.  

\emph{Technical details:} A deep dive into the system's inner workings, such as its architecture, data sources, goals, performance metrics, validation, and improvement history. 

\emph{Legal and ethical considerations:} Identification of legal, ethical, and other RAI-related concerns and ascertaining whether and how they have been addressed. 

\emph{Stakeholder analysis:} A critical dimension of the case study involves identifying and comprehensively analyzing diverse stakeholders associated with the AI system. This includes surfacing stakeholder perspectives and goals, examining their level of participation in the design, development, evaluation, and oversight of the system (if any), and assessing the benefits and harms for each stakeholder. 

\emph{Transparency and explainability:} A closer investigation of these two specific RAI principles, focusing on their relevance to the goals and perspectives of different stakeholders. 

\paragraph{Stakeholder matrices.} Next, we adopted a structured assessment approach to create comprehensive matrices that systematically outlined the benefits, harms (and their possible origins), tensions, and strategies for tension mitigation or resolution between stakeholders. We filled out these matrices for each case study and used them to guide the design of interactive workshop activities with facilitators, using tailored questions to prompt and structure discussions in breakout groups. In these matrices and all other training materials, we intentionally avoided formal definitions of RAI concepts like fairness, agency, and safety. This approach shifted the focus from terminology to encouraging participants to interpret these concepts within specific case studies.

To encourage discourse, we adopted an interactive learning model centered on matrix structures as frameworks for workshop participants (see Figure \ref{fig:jamboard_day1}). These matrices featured placeholders for learners to complete, with facilitators guiding discussions using handouts. By providing the matrix structures without pre-filled content, we empowered practitioners to become co-creators, filling in the matrices with insights, ethical considerations, and real-world examples from the case studies. This hands-on approach ensured practitioners were active participants in developing their understanding of RAI, rather than passive recipients of information. By placing learners at the center of knowledge construction, this approach transcended traditional teaching methods,  empowered them to actively shape their understanding of AI ethics and responsible decision-making through experiential and interactive learning.

\paragraph{Handouts.} Next, based on case study documentation and matrices, we crafted  practitioner-friendly handouts for the workshop. These handouts distilled the key elements of each case study into a clear, concise, and jargon-free format, providing participants with a structured guide to navigate the complexities of the cases. Practical case studies were presented as short stories, including excerpts from newspapers or magazines. To encourage active engagement and critical thinking, the handouts included discussion prompts and provocative or controversial statements. 

To enhance the training approach, we created two versions of each handout: one presenting the technical system in a positive light and the other in a negative light, particularly in terms of its repercussions for affected individuals (see Figure~\ref{fig:housing_handout_negative} for an example). This strategy fostered critical thinking and promoted a comprehensive understanding of RAI.

\section{Workshop implementation}
\label{sec:workshop}

\paragraph{Learning outcomes}

The learning outcomes of the workshop are grouped into three categories: (1) RAI concepts, (2) stakeholders, benefits, and risks, and (3) risk mitigation strategies. Upon completion, learners should be able to: Identify and define basic RAI concepts; Recognize RAI concepts relevant to a specific system, product, or service (``system'' for short); Explain RAI concepts related to a specific system to their team members; Identify key \emph{stakeholders} of a specific system; Identify the \emph{benefits} and \emph{risks of harm} associated with a system for each stakeholder, and relate these to \emph{RAI concepts}; Identify tensions between the benefits and risks of a system across stakeholders and apply techniques to reconcile these tensions; Explain stakeholders, benefits, risks, and tensions to members of their team; Analyze how specific harms may arise in relation to a system and how its data, technical properties, and context of use may increase these risks; Propose and describe \emph{mitigation strategies} to address specific risks of a system; and Describe potential harms and mitigation strategies to their team.

\paragraph{Workshop schedule.}  The workshop was iteratively designed based on suggestions from pilot participants and conducted in two 120-minute sessions to minimize participant attrition. To enhance interactivity, participants spent 50\% of the time (60 minutes per session) actively working through the case studies in small moderated groups. The workshop was offered twice in quick succession to accommodate all interested participants while keeping group sizes small enough to ensure active engagement.  Session 1  was structured as a sequence of six activities: 
\begin{itemize}
\item Introduction and welcome (10 min);
\item Lecture (20 min);
\item Moderated case study discussion in break-outs  (30 min);
\item Report-back to the full group (5 min);
\item Moderated case study discussion in break-outs  (30 min); 
\item Closing reflections (10 min).
\end{itemize}

Participants in different break-out rooms worked through different versions (positive/negative) of the same case study. Discussion during the first 30-minute activity revolved around stakeholder identification, goals and priorities, and benefits and harms to a selected set of stakeholders.  The second activity continued with the same case study and focused on using RAI concepts to identify and reconcile tensions.
Session 2 followed a similar structure. During Sessions 1 and 2, we presented concrete examples to illustrate RAI concepts, providing essential context. This initial exposure enabled all participants, including those with limited prior experience in RAI, to meaningfully engage with these topics in group discussions. The first case study was pre-defined and examined reasons for harm and potential socio-technical mitigation strategies. The second case study was open-ended, inviting participants to compose and describe their own scenarios, identify stakeholders, and analyze and reconcile the benefits and harms for different stakeholders.

\paragraph{Selected case studies.} We selected two case studies for the workshops.  We used the \emph{housing ads delivery} case study to support Session 1 activities, see Figure~\ref{fig:jamboard_day1} for a snapshot.\footnote{Additional details about the case studies are available in the full version of the paper~\cite{this_arxiv}.}  Participants explored the benefits of personalized ad delivery for vendors, advertisers,  advertising platforms, and customers. They juxtaposed these benefits against the harms of bias and discrimination against historically disadvantaged groups, as well as the loss of agency and control experienced by individuals targeted by the ads. 

We used the \emph{toxic content moderation} case study during Session 2. Participants considered automated moderation systems that flag and remove comments identified as toxic using supervised machine learning methods. These systems may be prone to bias, including pre-existing bias (e.g., if training data contains a disproportionate number of toxic comments from a specific demographic group, leading the system to flag comments from this group more frequently) and technical bias (e.g., if the system prioritizes certain types of comments during training or is tuned to favor specific performance metrics). Participants discussed how toxic comment moderation can help users reclaim their virtual space after being victims of online bullying.  They also discussed scenarios where social media platforms fail to flag toxic comments deemed ``not harmful enough'' by their guidelines, or erroneously flag and remove valid posts and supportive comments.

The dual (positive vs. negative sentiment) prompts for both case studies stimulated participants to think critically about the presented situations, boosting deeper discussion by adopting disparate perspectives. 

\paragraph{Practitioners as case study creators and assessors.} At the end of the educational process, practitioners become creators and assessors of case studies informed by their personal and professional experience. This activity involved designing real-world AI scenarios, describing  systems with controversial effects on diverse stakeholders, creating assessment matrices, identifying benefits, harms, tensions, and mitigation strategies, and conducting peer reviews. This method empowered practitioners to apply RAI principles in practice and simulate the assessment of real-world case studies they may design, develop, or deploy in their daily work.

\section{Analysis of Engagement and Learning}
\label{sec:analysis}

\paragraph{What was successful and what were some challenges?}  We observed that workshop participants stayed interested and engaged throughout the sessions. We attribute this sustained engagement to the relevance of the case studies and the ample opportunities to actively engage with the material. The quality of the interactions was remarkable, with participants contributing diverse perspectives from different stakeholder viewpoints and proposing effective strategies to mitigate potential tensions. This engagement was most evident when participants created thorough case studies from scratch, demonstrating both active participation and satisfactory level of understanding. Unfortunately, attrition remained a challenge: with just over 50\% of the participants from the first session returning for the second.

In the remainder of this section, we analyze  engagement and learning
based on workshop observation and pre-/post-survey results. We note that the cohorts across the two workshop offerings represent a range of technical and non-technical roles (see Table~\ref{tab:roles}). 

\subsection{Measuring engagement}
\label{sec:analysis:engagement}
Participants reported feeling ``curious,'' ``excited,'' and ``hopeful'' about workshop content.  They cited rationales for participating such as having a limited background on AI and RAI topics, recognizing the positive and negative transformative potentials of AI, and wanting to examine case studies and strategies for deploying AI  responsibly.  These adjectives align with the majority of participants (71\%) lacking specific expectations for the workshop, indicating they were receptive to learning more without preconceived agendas. Over 44\% of participants had never previously worked with RAI concepts, while the 29\%  who had familiarity were primarily interested in gaining a general understanding about RAI.  

We compare our sample with responses from the 2019 Mozilla Foundation survey on public perception of AI~\cite{mozilla}.  Among the 67,000 respondents in that survey, the most common adjectives describing AI were curious (30\%), hopeful (27\%), and concerned (32\%).  In contrast, responses from our workshop participants were more positive and less concerned. This difference is likely because our sample consists of technology company employees, who may be inclined to view AI in a more positive light. 

Among the participants across the two workshops, 29 (76\%) viewed RAI as important and 5 (13\%) viewed it as somewhat important to their day-to-day work. Among the remaining participants, 2 (5\%) were unsure about the connection of RAI to their work, and the remaining 2 (5\%) responded that RAI is not important to their work.  

Further, an average of 61\% of participants thought that AI will definitely improve their life, 37\% thought it might improve their life, and a single participant thought that AI would not improve their life. We note a 16\% discrepancy between the perceived work-relatedness of RAI and the notion that AI will improve life. In comparison, the 2019 Mozilla Foundation survey found that 24\% of respondents believed AI would make society better, 10\% remarked that AI would make society worse, and 41\% voiced a nuanced view, recognizing both positive and negative impacts.  This difference likely reflects our sample's AI-optimistic  perspective, as the participants were practitioners at a technology company, in contrast to the general public polled by Mozilla. 

A key aspect of the workshop was engagement with the provided case studies.  In the post-workshop survey, respondents ($n = 16$) reported that the most important stakeholders to consider are current or potential users and society at large, indicating that participants were examining RAI beyond their immediate work-specific contexts. Most respondents (77\%) reported being engaged by the participant-to-participant interaction provided through small case study discussion groups and hands-on activities. Additionally, 100\% of respondents agreed that the case study format was an effective for understanding RAI concepts.  

\begin{table}[t!]
\centering
\small 
\caption{Workshop participant roles.}
\begin{tabular}{ l r} 
 \textbf{role} & \textbf{number of participants} \\ 
 \hline
 content / media specialist & 3 \\
 designer & 10 \\
 engineer / research scientist  & 12 \\
 legal / policy adviser & 4 \\
  project manager & 9 \\
 \hline
 \textbf{total} &
 \textbf{38} \\
 \hline
\end{tabular}
\label{tab:roles}
\end{table}

\subsection{Measuring learning about RAI}
\label{sec:analysis:learning}

The case studies formed the core of workshop interaction. We used Jamboard, a free visual brainstorming tool by Google, to facilitate simultaneous group activities.

\paragraph{Session 1.} Figure~\ref{fig:jamboard_day1} shows a representative sample of the Jamboards.  In the left-hand image, one group---similar to other groups---successfully identifies stakeholders in \emph{housing ads delivery} who may be affected by the use of AI in this case study.  They also identify a wide range of positive and negative impacts of this systems on stakeholders. Across groups, practitioners focused extensively on the system's implications for individual users and the wider public, including issues of fairness, privacy, and safety.  Survey results indicate that practitioners resonated positively with the case study. The generated ideas and dialogue indicate that the case study effectively prompted thinking about the complexities of RAI, including competing stakeholder interests and strategic interventions. 

Part of Session 1 involved ranking the benefits and harms as primary and secondary impacts, linking them to RAI concepts, and recommending mitigation strategies. Interestingly, some groups dedicated time to creating a taxonomy to distinguish primary vs. secondary concerns. While certain groups focused primarily on the benefits or harms to specific stakeholders (\eg technology vendors or users), others approached the task by considering societal impact as the  basis for the  primary vs. secondary distinctions. 

We observed that, across groups, users elicited the most primary harms, while the platform elicited the most secondary harms.
Participants indicated no difficulty in adopting the user's perspective and recognized the platform's ultimate goal as a user-facing system.  Participants discussed how to interpret RAI concepts like privacy, robustness, fairness, inclusion, or accountability with respect to stakeholders, and, further, how to  map RAI concepts to benefits and harms.  Mitigation strategies proposed by participants included technical, policy, and design interventions, though they consistently reported feeling they had insufficient time to explore these actions in depth.

\paragraph{Session 2.} This session focuses on applying RAI principles using \textit{toxic comment moderation} case study. Participants found this case study to be both realistic and positively challenging, showcasing the difficulties in addressing RAI concepts. Many supported the use of authentic case studies, with one participant stating, ``the more applied the case studies are, the more valuable the workshop would be in the future.''  Practitioners focused on harm mitigation strategies as they relate to RAI concepts.  

By design, the \textit{toxic comment moderation} case study surfaced more complex considerations than the \emph{housing ads delivery} case. When focusing users as stakeholders, several groups noted that users are not a monolithic group, making harm mitigation strategies under RAI principles non-uniform. Case studies highlighted  different ways of perceiving technology, providing, as one participant described, ``real-life positive and negative examples of AI (e.g., Amazon's AI recruiting model that was biased against women)'' (in reference to~\cite{amazon}). 

The final interactive activity challenged participants to develop an original case study based on their personal or professional experience.  The topics covered a diverse range and included (1) classification of insurance claims, (2) autonomous vehicles, (3) job advertisement targeting systems, (4) social media beauty influencers, and (5) virtual reality characters. Participants increasingly utilized RAI concepts and terminology compared to the beginning of the workshop, as evidenced by dialogue and content analysis.  Additionally, they began discussing tensions more quickly than during the initial breakout session. 

\begin{figure}
    \centering
    \includegraphics[width=\linewidth]{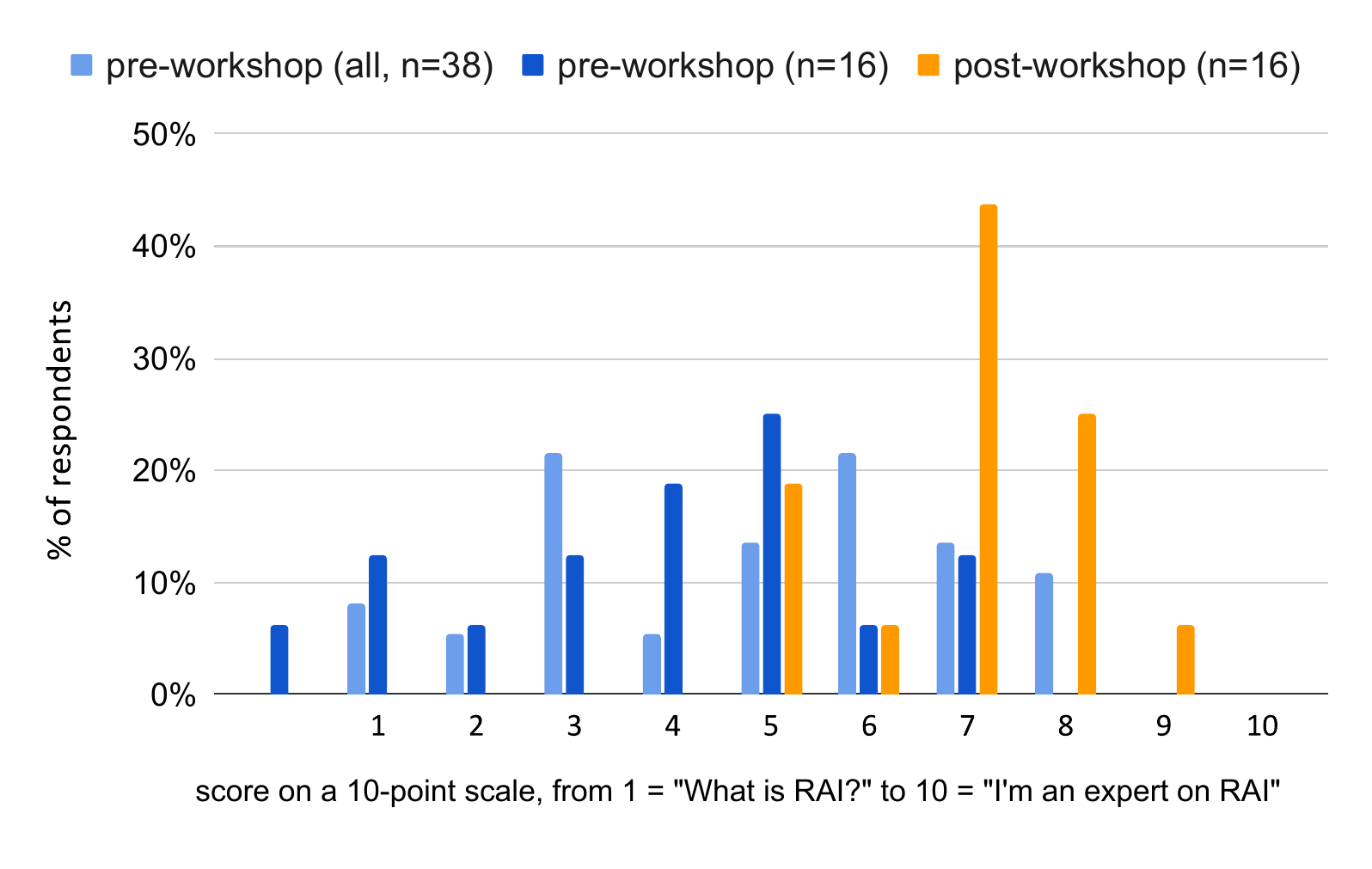}
    \caption{Self-assessment of RAI knowledge: pre-workshop responses from all participants ($n=38)$ in light blue, pre-workshop responses from participants who also submitted post-workshop surveys ($n=16$) in dark blue, and post-workshop responses from the same participants in orange.}
    \label{fig:prepost}
\end{figure}

\paragraph{Assessment of implementation.}

We conducted pre-/post-workshop self-assessments of RAI knowledge, with the results summarized in Figure~\ref{fig:prepost}.  Of the practitioners who participated, 38 completed the pre-workshop survey, and 16 completed both the pre- and post-workshop surveys.  Participants rated their post-workshop understanding of RAI concepts and techniques significantly higher, on average, than their pre-workshop understanding: a weighted average of 4.88 (for all participants, $n=38$) or 4.84 (for participants who submitted both surveys, $n=16$) versus 6.94. All participants responded that their knowledge of RAI concepts improved at least somewhat, with 77\% reporting that their knowledge was very improved to substantially improved, and 100\% reporting that they are motivated to apply workshop content to their own work.  Participants appreciated the workshop's compact format---two 120-minute sessions---but also expressed interest in exploring the material further and continuing their study of RAI.  

\section{Conclusion}
\label{sec:conc}

In this paper, we discussed our experience teaching Responsible AI (RAI) to industry practitioners at Meta. Our assessment results indicate that participants, many with limited prior knowledge, found the workshops engaging and reported a positive shift in their understanding of RAI, and in their motivation to apply it to their work. Using case studies relevant to participants' work activities was particularly well-received. As a further development, we are preparing a public website to share our workshop structure and case studies as a collaborative educational resource.  Below, we outline what we believe are the best practices for crafting case studies to teach RAI effectively to practitioners.

\paragraph{Case study selection.}
Begin by carefully considering your practitioner audience's background, knowledge domains, interests, and goals. Tailor case studies to align with their specific needs and objectives. Use case studies most adherent to the participants' daily activities or organizational reality.  Further, opt for case studies that offer depth and complexity, encompassing a range of intricacies, challenges, and ethical dimensions to support a holistic exploration of RAI.  Finally, prioritize scenarios that involve multiple stakeholders with divergent perspectives, as these tensions encourage learners to navigate the complexities of AI decision-making in real-world contexts. For breakout room activities, structure groups to be heterogeneous, mixing participants with diverse backgrounds, roles, and perspectives on AI and RAI.

\paragraph{Case study materials.}  Develop matrices that systematically outline benefits, harms, tensions, and strategies for tension mitigation or resolution across stakeholders. These matrices serve as tools for interactive learning, helping participants deepen their understanding and encouraging the adoption of RAI practices.  Further, craft two or more versions of handouts. In our case, we developed one with a positive and another with a negative portrayal of the AI system's impact. This multiplicity prompts nuanced exploration and insights into framing effects, fostering diverse perspectives into the discussion and prompting participants to reflect critically on their own viewpoints. 

\paragraph{Practitioners as case study creators.} Culminate the learning journey by empowering practitioners to become creators of case studies from scratch. This final stage fosters critical thinking, practical application, and a deeper understanding of RAI principles. Furthermore, it instills confidence in participants, ensuring they feel capable of applying the same process to real-world AI systems. Finally, this practice equips practitioners to effectively communicate RAI concepts to coworkers, navigate tensions, and facilitate future educational activities. 

\paragraph{Limitations and future directions.}  Our study is an experience report: it is limited to several workshops conducted for practitioners at a single---albeit one of the most significant---institutional partners in the industry.  This comprehension will be enhanced by feedback collected through the public website we plan to launch in the immediate future, which will host open educational resources for practitioner training. Despite the limited audience, the level of engagement, direct feedback, and participants' understanding of RAI concepts, as evidenced by their collective discussions, supports the use of case studies to advance deeper understanding of complex topics.

An area for further analysis is the longitudinal impact of case studies. Due to constraints in practitioner schedules and organizational partner's priorities, we were unable to provide extended programming, but we recognize the need to explore the impact of case studies on longer-term understanding of RAI concepts.  For example, one question of interest is how practitioners continue to engage with RAI concepts, both implicitly and explicitly, months after the workshop.  Another strand to examine is the ability to mitigate RAI threats, a topic that several participants identified as a natural follow-up to the workshop.

\section*{Acknowledgements}
\label{sec:ack}

The authors of this paper extend their gratitude to the dedicated members of Meta's Responsible AI team and other colleagues for their invaluable partnership in developing educational materials, case studies, and implementing workshops.   We are especially grateful to Parisa Assar, Emily McReynolds, Jacquieline Pan, Hunter Goldman, Eleonora Presani, and Miranda Bogen.  We are also grateful to Lucius Bynum, Venetia Pliatsika, and Lucas Rosenblatt from the NYU Center for Responsible AI, who co-facilitated the workshops.

\begin{links}
    \link{Extended version}{https://arxiv.org/abs/2407.14686}
\end{links}

\newpage

\bibliography{references}

\newpage

\onecolumn
\section{Supplementary Materials}
\label{sec:suppl}

In this section, we include the handouts for the positive and negative versions of the \emph{Housing Ads Delivery} and \emph{Toxic Comment Moderation} case studies. 

The \emph{Housing Ads} case study handouts are color-coded to highlight the stakeholders (in yellow) and the positive (green) and negative (red) sentiment sentences.  This color coding is given here for illustration only and was not part of participants' handouts during the workshops.

The \emph{Toxic Comments Moderation} handouts are not color-coded because their sentiment is due to the inclusion of an example on the second page (positive user experience for Alex, negative experience for Tina).  Both handouts contain an explanation of some of the inner workings of the toxic comment moderation system and of the types of bias that may be introduced.

Note that the reference to Google (\emph{Housing Ads}) and Twitter (\emph{Toxic Comments}) does not point to the identity of our \emph{organizational partner}.  The examples used here are generic and refer to well-known platforms that offer the services discussed in the case studies.

Finally, we include a snapshot of the Jamboard created by workshop participants during the final break-out session (Day 2 / Activity 3), where they co-constructed a case study around Automated Cars, identified stakeholders, RAI concepts and concerns, and explored tensions.

\newpage

\includepdf[pages=-]{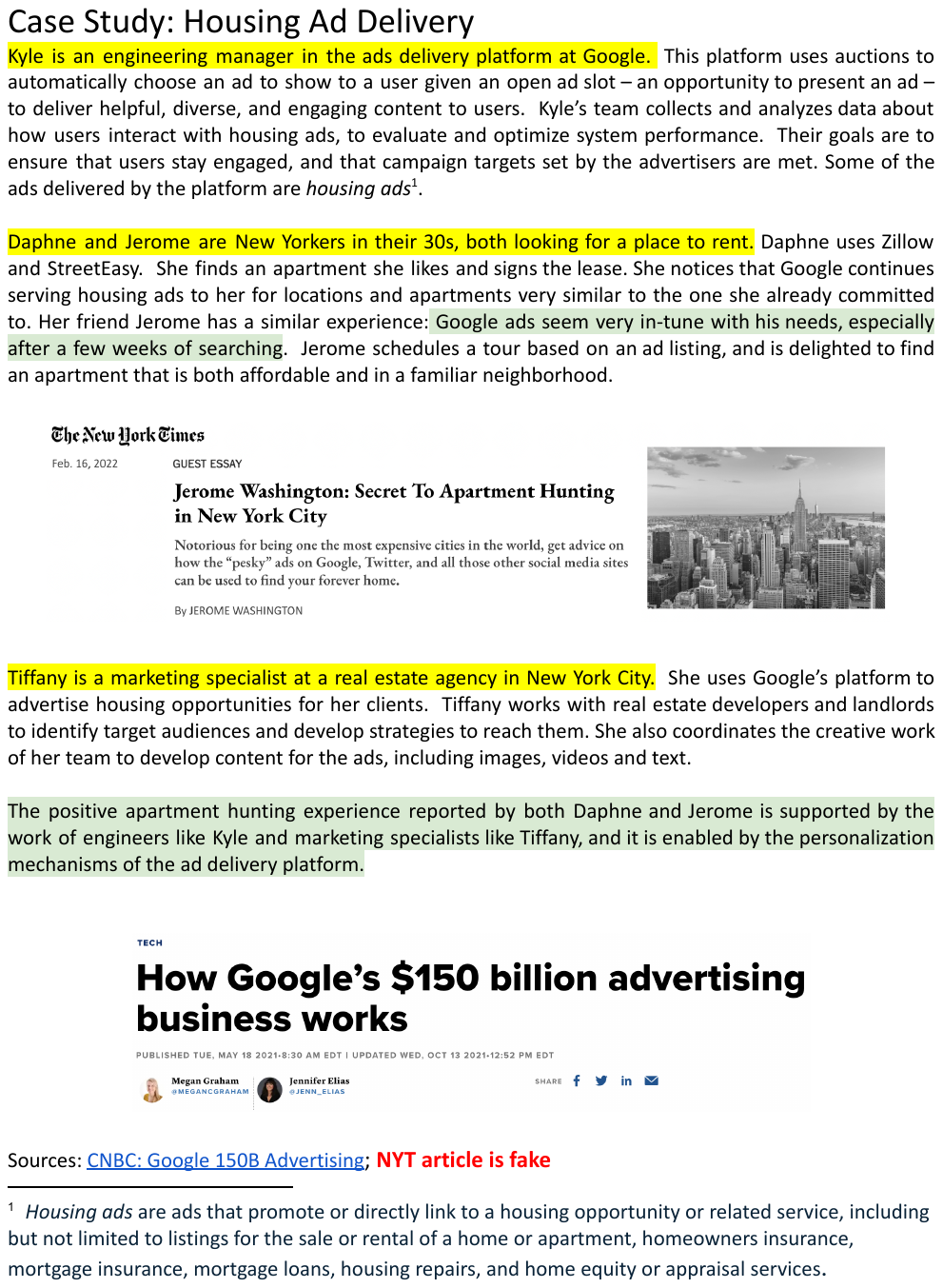}
\includepdf[pages=-]{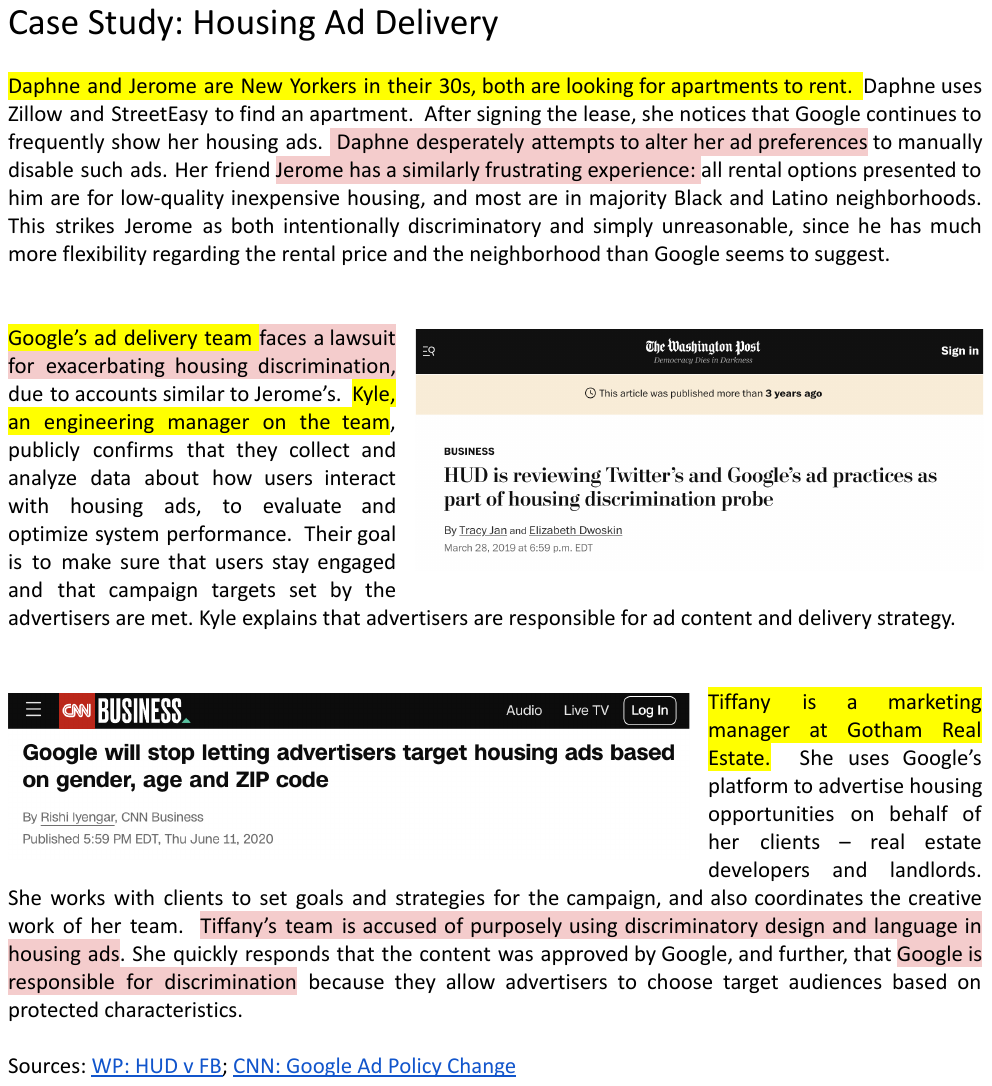}
\includepdf[pages=-]{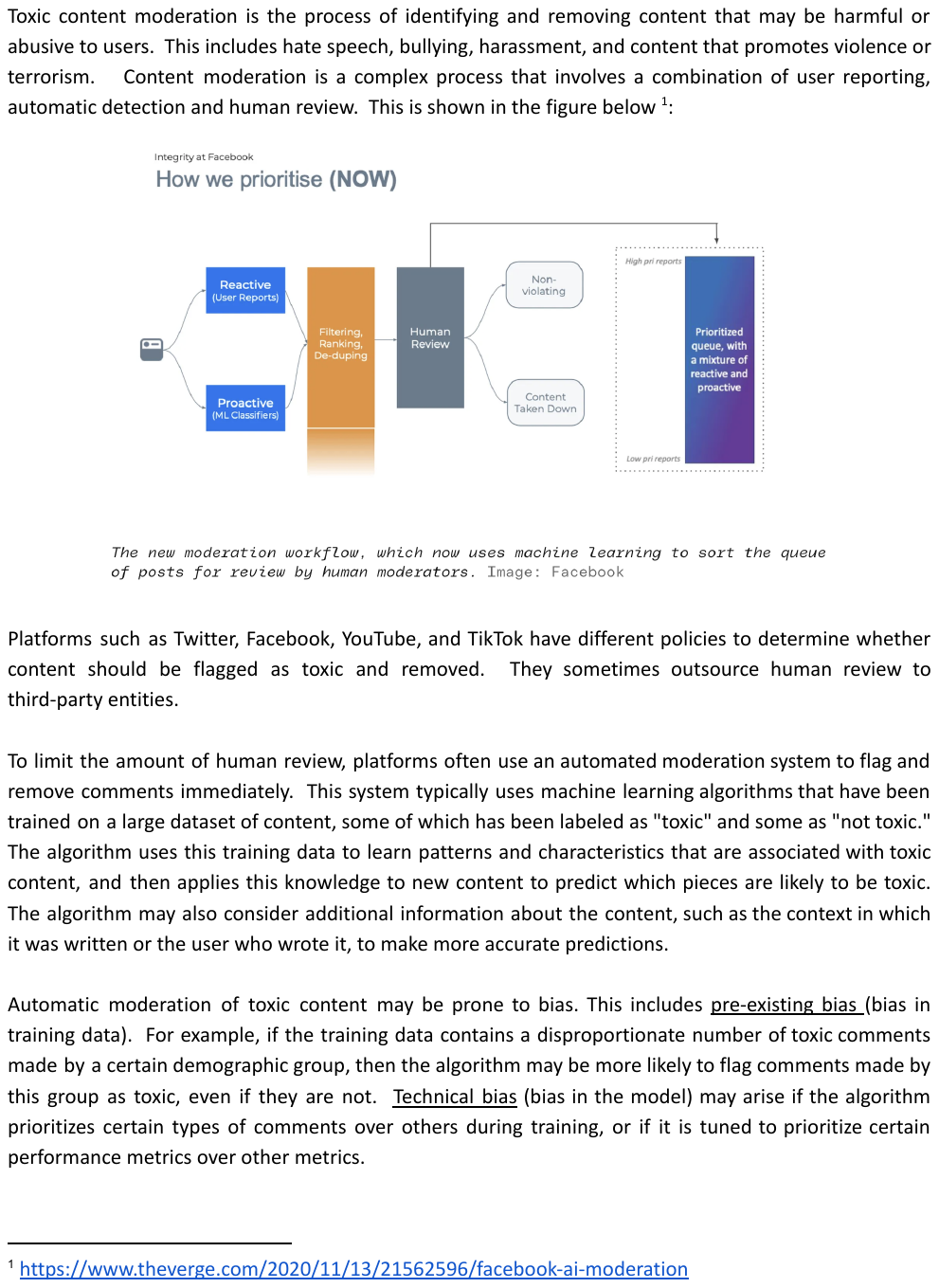}
\includepdf[pages=-]{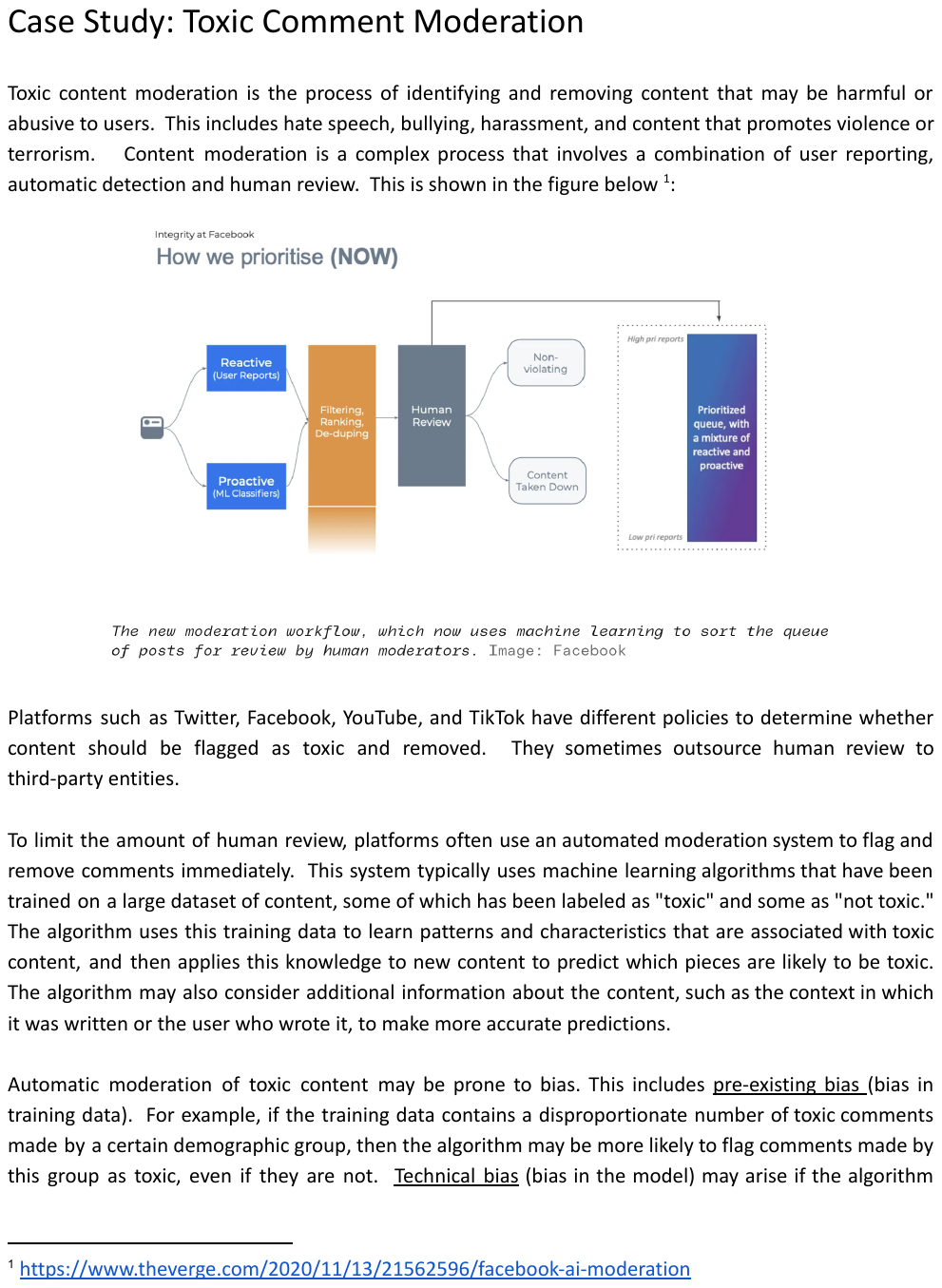}
\includepdf[pages=-]{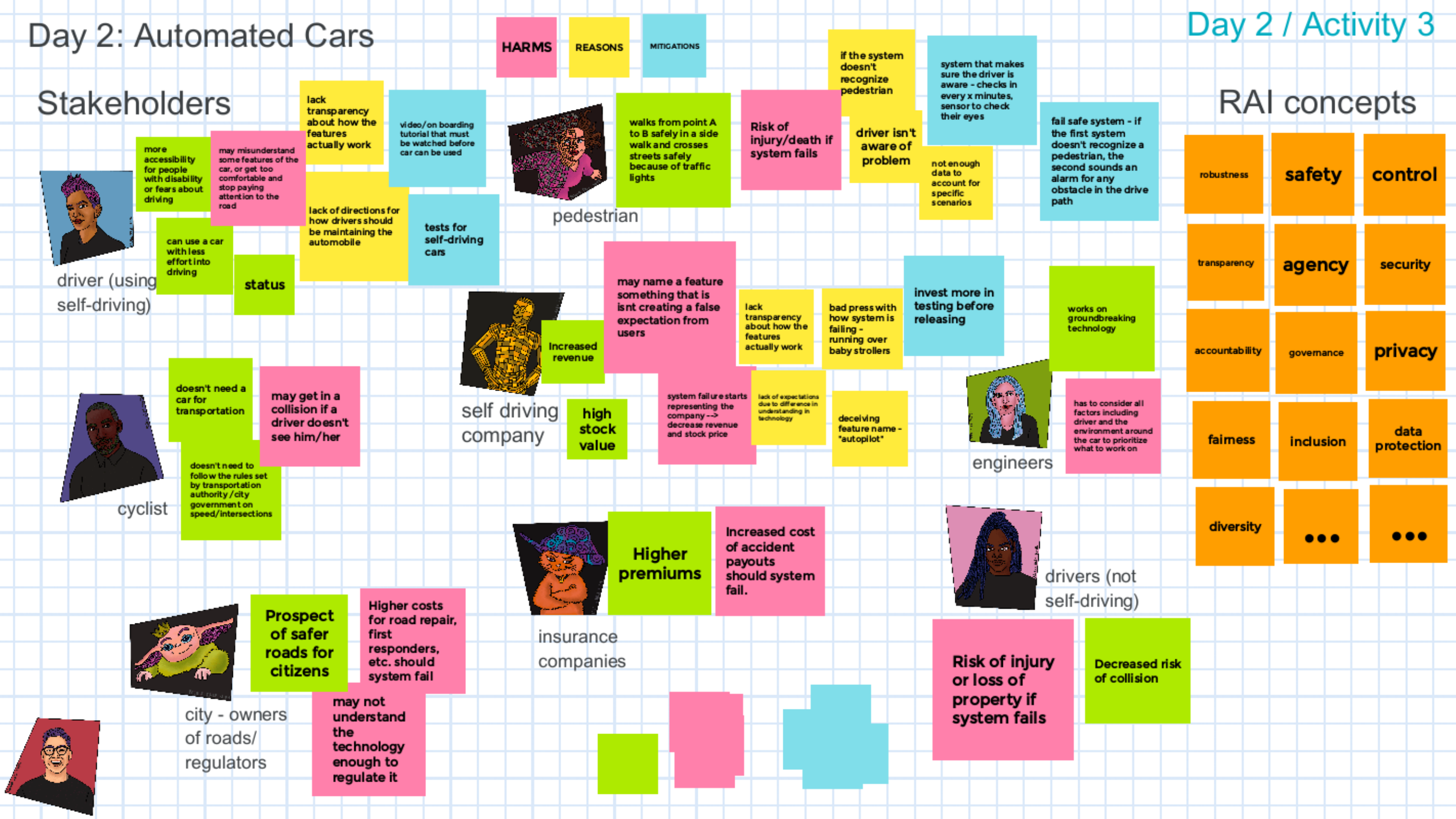}

\end{document}